# Oracle R12 E-Business Suite – Role Based Access Control and Roles Lifecycle Management

Sajid Rahim
Department of Computer Science/Software Engineering
McMaster University
Hamilton, Ontario, Canada
rahims9@mcmaster.ca

*Abstract*—Oracle E-Business Suite R12 is a widely used ERP solution that provides integrated view of information across multiple functions and sources. It allows for simplified business process tools for Shared service model e.g. Centralized Operation where multiple operating units can be supported. Security considerations are vital for such operations in large enterprises. R12 introduced Role Based Access Control security based on ANSI RBAC standard. R12 RBAC implementation is challenged with lack of Roles Lifecycle Management (RLM) process which also contributes to challenges such as Segregation of duty (SOD), and controlling access to PII for multi-country operation for common functional areas. The paper will propose a possible Roles Lifecycle Management process.

## I. Introduction

Businesses are continuing to implement mission critical applications such as Oracle E-Business Suite R12 or Fusion by replacing legacy and home grown systems with the aim of standardized business processes and IT infrastructure while providing single view on the business. Business users are granted access based on their roles as defined by Human Resources.

E-Business Suite R12 has embraced Role Based Access Control (RBAC) based on RBAC ANSI standard (ANSI INCITS 359-2004) originally proposed by the US National Institute of Standards & Technology (NIST). RBAC provides secure controls for access and authorization that can be used to create and enforce segregation of duties. This co-exists with R12's 'traditional' responsibility based access control. R12 RBAC layers upon responsibility access control as Roles. Roles provide significant security design improvement over the Responsibilities based option by normalizing access to functions and data through user roles rather than only users.

Since R12 RBAC layer is relatively new, there is a lack of Roles Lifecycle Management (RLM) process. RLM is an encompassing process by which an organization develops, defines, enforces and maintains roles access control. Effective RLM is essential for improved security controls for access and Segregation of Duties (SOD) which assist with audit and control compliancy. Role based access control is not a one type project as roles are dynamic and must evolve as business changes. Effective role management requires a continuous process for monitoring and managing.

The paper is motivated in proposing a Roles Lifecycle Management process which can be used for effective controls and audits for R12 EBS as a must have in order to assist with providing insights into which user has access to what.

## II. Background

### A. Roles and Definition

Role Based Access Control (RBAC) is as widely used access control method which assigns users to Permissions through Roles (User-Roles relationship). Roles are based on job functions and can be derived from organization roles. New roles can be derived though hierarchy though role inheritance (Role-Role relationship) or changes in permissions (Roles-Permissions relationship). These components collectively determine as a role member, what user constraints are in terms of what access or operation that may be performed on the data/information or transactions. In other words, it presents an abstraction of organizational responsibilities as a role and its relationship to the organization which is assigned to respectful users.

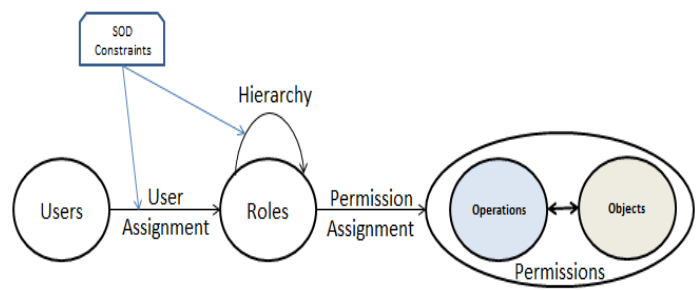

Fig 1. Role Based Access Security

Roles provide significant advantages in terms of compliance including segregation of duties, administration of users, secure and need to purpose based access control. Additional RBAC derivatives are now available such as EnRBAC,Temporal RBAC.

*B. Oracle RBAC Security*

Traditional 'Responsibility' provides authorization to a user with a list of function security, Fig2. Two types of functions exist namely executable and non-executable. An executable function is invoked from the menu or submenu page which is presented upon user login. The function will invoke a transaction, page, report or service. A non-executable function is an abstract alias which can be used to define a submenu, a subset of a page or form. A menu is created by grouping a set of functions in a hierarchical structure with submenus, prompts[5].

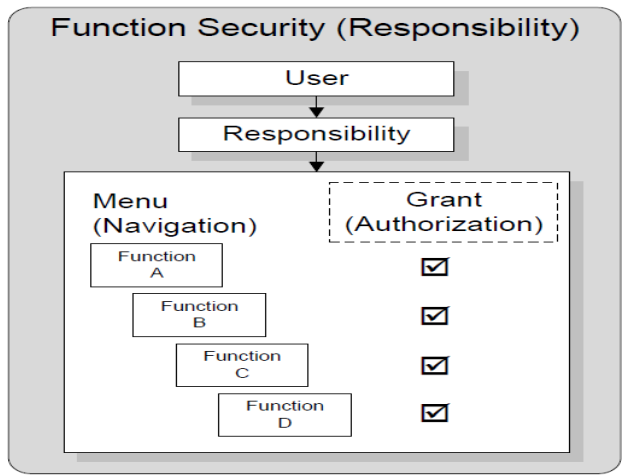

**Fig 2. Function Security**

A responsibility provides authorization to a menu of functions based on a user role. A user can have one or more responsibility. Many users can share the same responsibility.

RBAC security shares the same concepts but expands the definition. Authorizations are made via Permissions at function level; grant authorization is discontinued. Typical RBAC provides access to objects in the form of permissions – in EBS R12 this becomes access to a function. These permissions are grouped together as grant permission set across accessible functions. Roles are assigned to users based on their roles in the organization and user performs only those functions are defined by the required role within the organization, Fig 3.

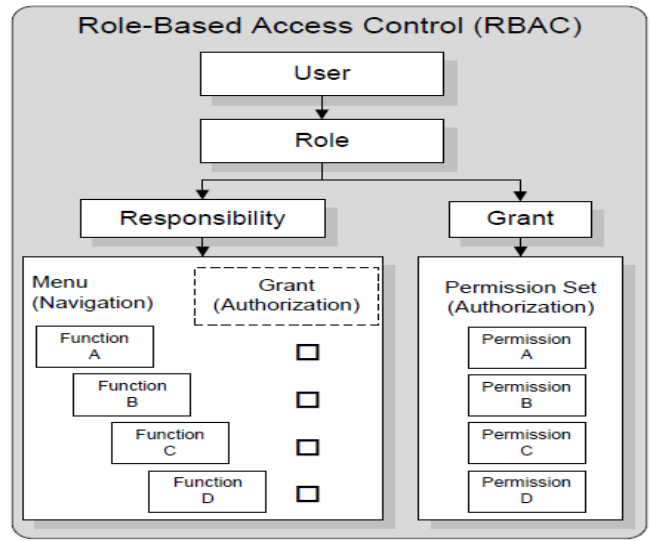

**Fig 3. Role Based Access Control**

For a single role, RBAC appears to be an over kill but a user in general has multiple responsibilities. Hence RBAC becomes very useful in creating a single role e.g. Sales Role that allows multiple responsibilities such as Employees and Sales as shown in Fig 4.

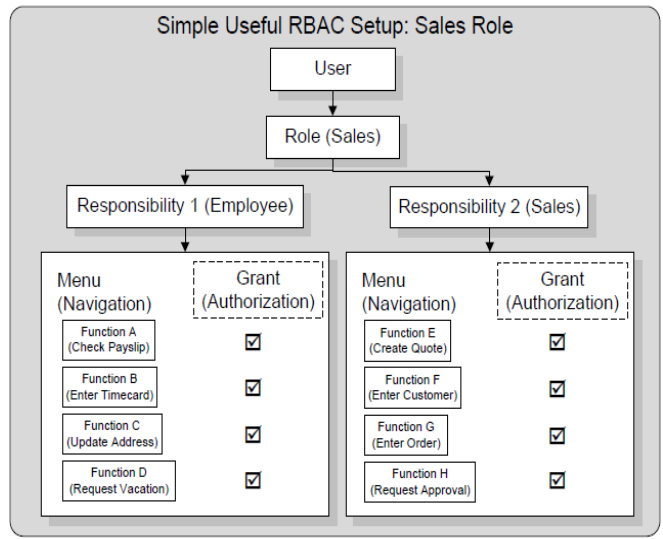

**Fig 4. Role defines multiple responsibilities**

A Role can be tuned with selective authorization via separate Grants and permission sets within same responsibility. For example, a sales role can be for Sales Person or Sales Manager; Sales Person can view the orders but Sales Manager can view and approve the orders.

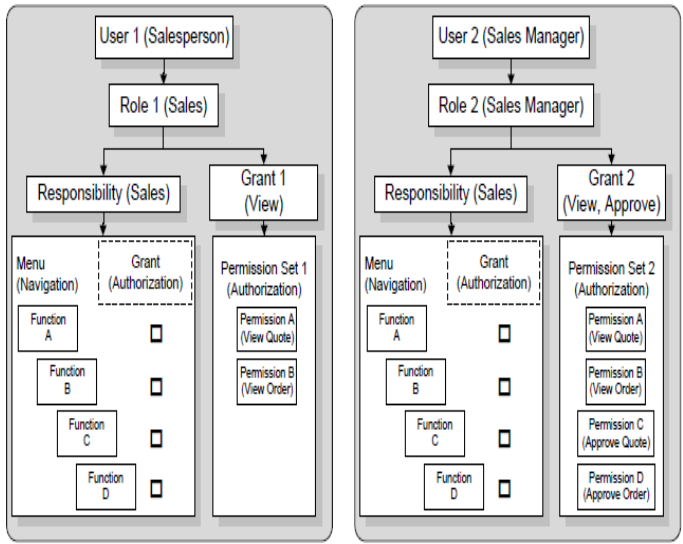

**Fig 5. Function Security**

Finally in R12 EBS RBAC, Roles hierarchy is comes into effect when roles inherit permissions of roles. Users are assigned roles once Permissions are assigned to these roles. In Fig 5, Manager inherit Buyer's role which inherit Inquiry Role. Purchasing Buyer is assigned Buyer role while Purchasing Manager is assigned Manager Role.

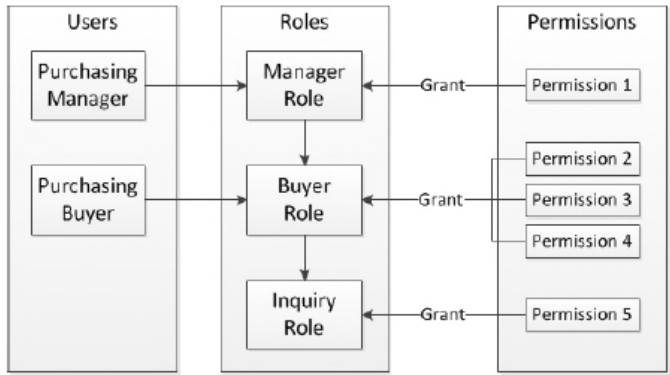

**Fig 6. Roles Hierachy**

Roles hierarchies further can be utilized to group together multiple responsibilities, and multiple roles together with various combinations of grants/permission sets, Fig 6.

To summarize R12 EBS RBAC provides offers better security and greater productivity through Roles whilst reducing overhead associated with traditional responsibility access control. Because of R12 EBS dynamic environment, it necessitates Roles Lifecycle Management to manage roles and associated users from inception to retirement.

*C. Oracle R12 EBS User Management.*

We note that Role Lifecycle Management is a necessary process that is required for successful maintenance of R12 EBS RBAC. Roles Lifecycle Management describes changes to Roles starting with the creation, modification, and deletion of roles proceeded by assigning and revoking of roles to users. It includes tracking these changes for audit purposes. The paper will now proceed on describing how existing components of R12 EBS can be leveraged to support a RLM process which will be described in the next section.

In R12 EBS, user security is offered by User Management functionality which has six built up layers as shown in Fig 7 [6]. Of interest are three core security layers starting with Function Security that describes what a user can do, Data Security describes what a user can see and Role Based Access Control layer which grant access to roles that include function/data security. Administrative Features allow for workflow based self-service and approvals including delegated administration of roles.

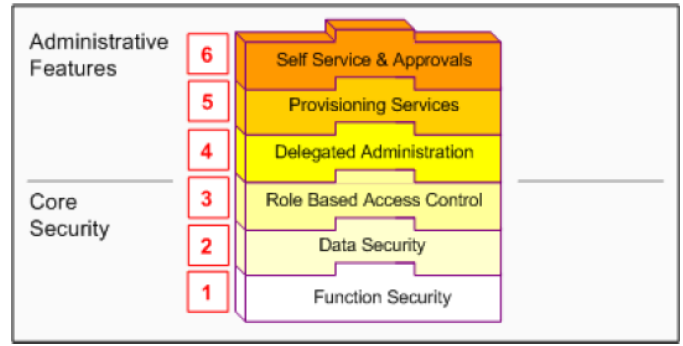

**Fig 7. User Management Layers – Core Security**

Role Based Access Control layer supports the mapping of user access control based upon a user role in the organization. Roles are grouping of all the responsibilities, lower level permissions (functions), permission sets, and data security rules that a user requires for performing a specific task. Role Categories allow Roles are organized into groups.

The three layers are assisted using web forms which are available via separate R12 EBS responsibilities namely User Management Fig 8, Functional Administrator Fig 9, Functional Developer Fig 10.

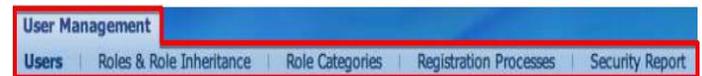

**Fig 8. User Management – Core Security**

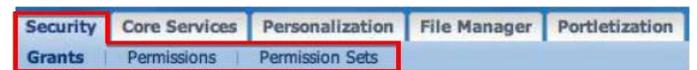

**Fig 9. Functional Administrator – Core Security**

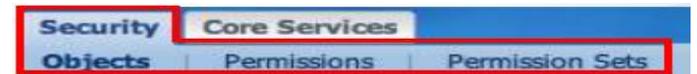

**Fig 10. Functional Developer – Core Security**

These sets of functional competencies allow for the creation of objects e.g. table/view which need to be secured using Functional Developer. Permissions which control access to functions and data, grouping of these permissions into a permission set, and giving grants for an action on a specified object to a grantee is managed by Functional Administrator. A grantee may be a role, group or a user.

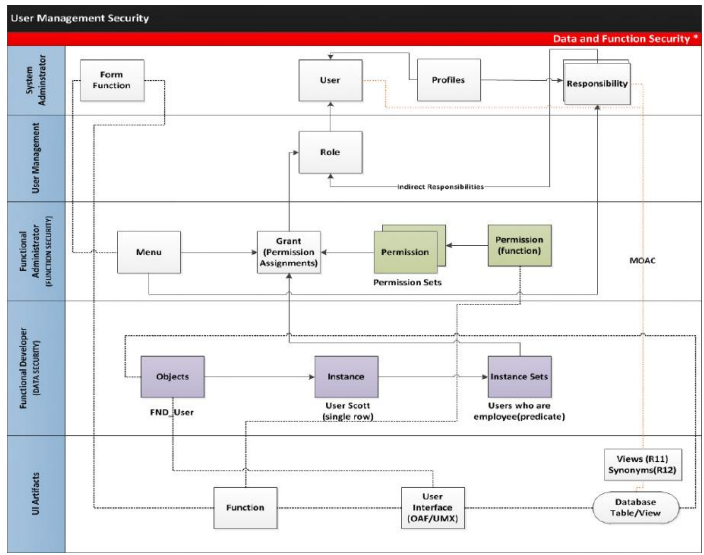

**Fig 11. User Management Security – Process Flow**

All the above shows a fairly robust set of RBAC associated functionality which can be used to create permission/permission sets, create object instance sets, create grants, and assign roles using User Management Security process flow, Fig 11. While there is a functional RBAC implementation, there is a lack of formal process by while Roles can be analyzed, defined, managed and maintained in the form of Roles Lifecycle Management.

## III. ROLES LIFECYCLE MANAGEMENT

### A. Present Methodologies

While RBAC is widely received and many research papers are available, very few papers describe what a lifecycle process will be. Both approaches described by Kerns et al.[7] and Schimpf [8] leverage software engineering methods. Additional role lifecycle such as process based and temporal roles lifecycle are unsuited for use within R12 EBS frameworks and therefore not discussed.

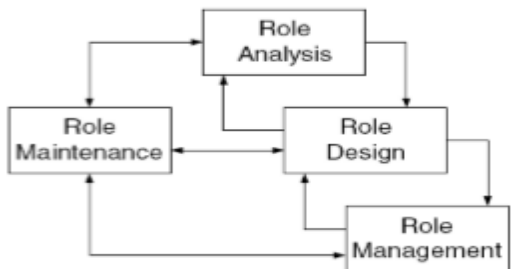

**Fig 12. Role Lifecycle Flow – Kerns et al.**

Fig 12 shows lifecycle consisting of four modules which are iterative in nature and without a strict sequence of events or details. Starting with Role Analysis which identifies roles within the environment context, Role Design transforms this definition into usable roles. Role Management focuses on daily administration and changes to the role model while Role Maintenance will deal with major changes in the roles due to organizational changes, user, roles and permission relationships.

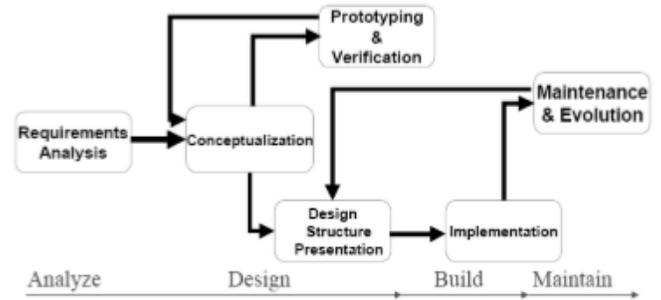

**Fig 13. Roles Lifecycle Management - Schimpf**

Schimpf [9], also follows software development lifecycle using Analysis, Design, Build and Maintain stages, Fig 13. Analysis defines the outcomes of the Roles Engineering Process. Design assumes a zero RBAC model presence and it seeks to define it while constructing roles prototypes and establishing Roles Catalogue. He moves to Build phase which implements the Roles definitions and finally Maintenance which manages changes that the organization will undergo.

Both models do provide a semblance of order with structure but they do not go into details of each phase or any interdependencies. Kerns fails to identify Role Catalogue which is a core component or documenting/capturing meta-data. Schimpf's model also fails to recognize that RBAC can be a standard feature within an enterprise system and focus needs to be on deriving roles and extending them.

This paper seeks to derive an improvised model after considering both of these two models which have commonality yet differences. Phases will be extended with additional phases and outcomes. Since a Role Model already exists within R12 EBS framework and the focus will be on the lifecycle alone that is easy to use and manage with.

### B. Proposed Methodology

Kerns et al (2002) first identified a cycle which is a natural fit for this use but it presents some limitations which we will attempt to extend and improve in specific context to Oracle R12 EBS RBAC. A role lifecycle can be adopted in the form of interative traditional waterfall. Four core components specified as Role Identification/Analysis, Role Definition/Enginering, Role Management and Role Maintenance/Governance. The flows noted allow for ease of movement between each stage.

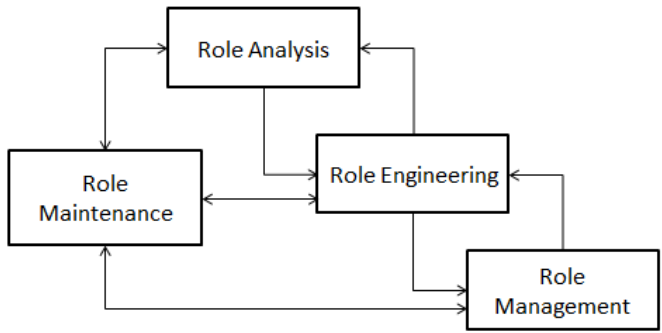

**Fig 14. Role Lifecycle Flow – Extended model**

An organization has separate divisions. This cyclical nature allows each of these processes to be performed incrementally for each of these divisions or operations. Should a challenge arise in the sequential stages whether it be a change in business process due to regulatory requirement or gap in requirements, the model allows to step back and fix prior stage and cascade aligned corrections down to subsequent stages. This iterative waterfall model presents advantages namely:

- Allows for earlier discovery, detection and remediation instead of finding problems later such as duplicate roles, unclear regulatory requirements, changes in business organization structures, changes in regulatory requirement due to external factors; this is critical to prevent redesiging entire organizational roles.

- This model allows for a pilot project be completed for one division. Lessons learned from the discovery process can lead to improvements in all the stages so that subsequent cycles for other divisions can be easily managed and even automated.

In summary, ontinuous life cycle is recommended as it guides towards more intelligent and efficient role design.

*D. 1. Role Analysis*

Role Analysis is the first core process of this particular methodology and our model provides core process details that will support this phase as illustrated in Fig 12.

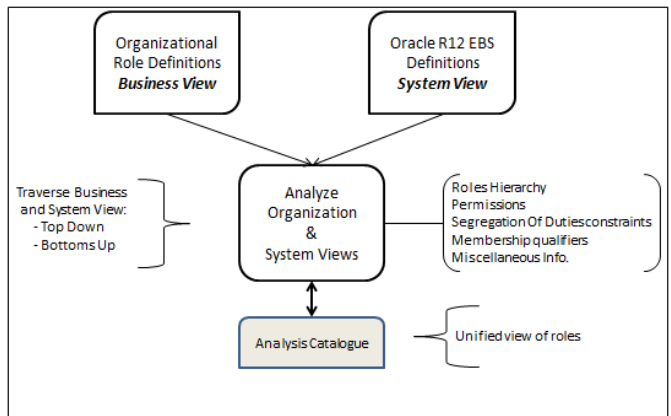

**Fig 12. Role Analysis Process.**

This phase as illustrated in involves Analyzing Organization and System views by consolidating of organizational structures, existing functional responsibilities, user groups, permissions, segregation of duties; these will represent Business Views while existing R12 EBS seeded roles definitions represent System View. These will analyzed by traversing the information in a bottom up and top down information review for maximized analysis. Role concept model, Roles Hierarchy, permission, segregation of duties constraints, membership qualifiers and other miscellaneous role support information will be derived. The results will be stored in Analysis Catalogue which will assist determine potential roles, permissions, and user/permission groupings. Analysis Catalogue will also assist with business validations and audit.

It is imperative to highlight that new roles are identified within the organization context using human resource and functional role description, duties and permissions. Following recommendations are noted for deriving sustainable roles.

- Roles need to be analyzed by functional specialists with explicit knowledge of the subject organizational roles.

- Leveraging formal ontology to describe jobs, tasks and access to EBS data/functions after creating an hierarchical structure of the organization.

- Adhere roles to natural structure of the organization as this tends to be generally static. Changes in structure will increase role management.

- Hybrid analysis of identifying all permissions that exist within the organization, and clustering these by traversing the organization structure either up or down to identify permission clusters which will constitute a role.

- Access control can be compromised by complexity. Higher levels of complexity can result in lower security as control cannot be defined. Roles must be defined with the permission cluster maps to natural role e.g. permission cluster will match a user community group.

The real value of roles is realised when they are analyzed to be able to operate a the level based on the access requirements for specific job functions or business processes; this assists elimininating compliance violation and mitigating risks.

*D. 2. Role Engineering*

This phase will build on the results of role analysis which has completed the organization review as well as standard roles which are delivered by R12 EBS. Meta data from the analysis will reside in Analysis Catalogue and used during this phase as well as for comparative purpose during the Roles Maintenance phases.

Roles Engineering will conceive formal roles definitions which will be prototyped and verified in an iterative manner until business needs are met in the respective sub-processes. Once confirmed, these roles will then be catalogued as formal Role Concept Models which will be utilized as input to Management phase. This phase is supported by the following processes namely Design Roles, Prototype Roles, Verify Roles leading to Design Signoff that sees the roles now saved in Role Concept Model.

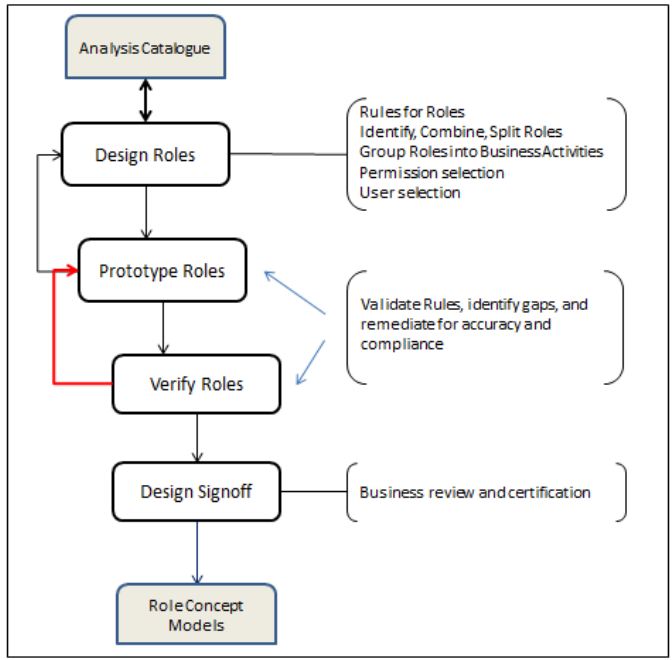

**Fig 12. Role Engineering Processes.**

Design Roles process will work in an iterative matter with the subsequent processes to refine the design. This may involve further qualifying rules for Roles concept models, further identification of new Role concept models including combining and splitting new Role concept models and grouping Roles by Business Activities, Permission Selection and User selection. Transient Roles concept models will be prototyped and verified.

Verify Roles process will be conducted jointly with business user; it will review Roles derived, memberships, permissions and user groups are valid; any changes will be iterated back with Design Role process, prototyped and reviewed thereafter until accepted by Design Signoff process which will catalogue an affirmative Role Concept Model that is tuned and tagged with Permission selections and candidate user selection lists. Analysis Catalogue will be updated by these findings. This collaboration is vital as it establishes a common language and introduces the data model to business users.

During the Design Role imperative to recognise that two main categories of roles are present. Roles that are conferred by the organization process and that which is administration of the roles themselves. The later in the case of R12 EBS will require the three functional roles e.g. Functional administrator, Functional Developer and User Management be made available for a group of administrators.

Following guidelines are noted during the Role Engineering process:

- Creation of a catalogue which maps each organization structure and functional role to an R12 EBS Role. Examine delivered R12 EBS Roles as re-engineering candidates as excellent source templates. The catalogue will capture Segregation of Duties (SOD) rules or dependencies for conflicting stored Roles.

- Multi-Organization Access Control (MOAC) simples security by allowing a single role to be used across various organization divisions.

- Create Permission and Permission Sets as needed User to Role association using Functional Administrator.

- Consideration for automated tools such as Desktop Administrator where roles can be engineered and business rules for roles including SOD are captured and conflict exceptions can be reported on.

- Identify metrics for measuring role quality e.g. is the role simplifying the access, is it well understood, how many people will use it, role constraints, future role expansion, compliancy violations etc and quantify these.

This is the most critical phase in this process and it is essential for functional specialists and Oracle R12 EBS Roles administrators to perform Conference Room Pilot (CRP) cycles to validate and confirm the roles correctly meet the business requirements. This aligns to Oracle EBS security model which recommends business owners review and approve Roles based on business needs, organization structure and user roles; this can also serve as a certification process.

D. *3.Role Management*

Role Management will comprise of two processes which work closely hand in hand; Role Deployment and Role Administration, Fig 13. Starting with Role Deployment that will focus on the implementing/deploying the Roles into Production R12 EPS Production/Test Environment as defined and signed off in Roles Concept Models from Roles Engineering phase. These implemented roles are stored within the application metadata and noted as Roles Catalogue. Once deployed, administering changes to deployed roles within the context of Roles Catalogue and Roles Concept Model as well as user administration becomes the domain of Role Administration process.

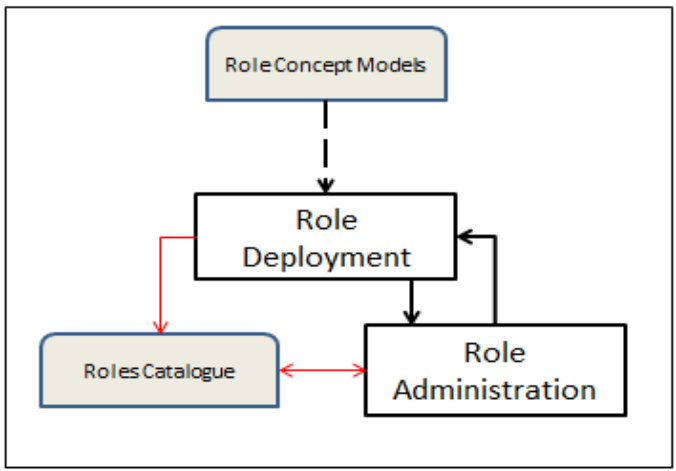

**Fig 13. Role Management sub-processes.**

Role Deployment, illustrated in Fig 14, is the implementation of the Role Concept Model into role templates that are created as custom Roles with additional permissions and security rules using a custom Wizard application which can be integrated with R12 EBS application; this new data is identified as Roles Catalogue. Users and user groups are assigned the appropriate roles from this Roles Catalogue. Metadata is ready for transfer and deployment using standard Oracle EBS fndload utility first to R12 EBS Test Environment for validation prior to repeating the process to R12 EBS Production Environment.

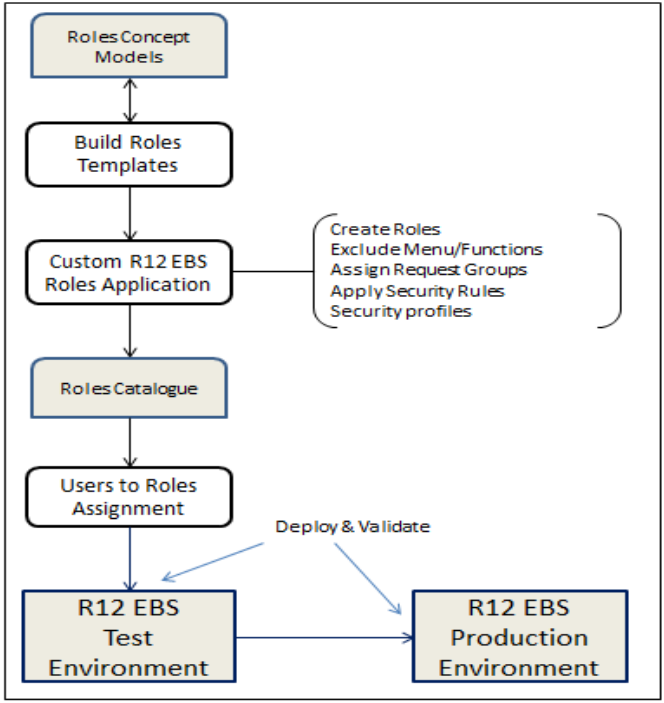

**Fig 14. Role Deployment process.**

As noted before Role Administration focus will be on designed roles that were catalogued from the design phase and deployed. Some of the activities will be:

- Organizational changes that impact existing definition of a role.
- Creation or deletion of user or permissions.
- Modifications including joining/removal of relationships between:
  - User/User groups – Role/Roles
  - Role/Roles – Permission/Permission Sets
- Roles changes which include:
  - Splitting a role into multiple roles due to Segregation of Duties etc.
  - Merging multiple roles into single role

Role Administration itself as defined can be well supported by period role check which can be partially automated; business changes affecting role definition can be preprocessed in order to identify the change which will affect the catalogued roles.

The above activities are all within the given constraint of an existing designed role. Any change related to an existing catalogued role will be handled in this area. This work is solely the domain of designated Roles administrators with Functional Administrator role.

### D. 4. Roles Maintenance

Roles Maintenance focuses on major changes to roles that have been catalogued for organizational use from prior stages. These changes originate from organization structure to role mapping, changes in user-roles and role-permission relationships.

Roles are never static and are bound to be subjected to re-engineering as organization structures or even organization will undergo change in the form or mergers, acquisitions or process re-engineering. Effective maintenance requires continuous monitoring and managing the maintenance to ensure effectiveness.

A Role Manager is required to adapt and revise the roles for the newer environment. It is logical for a method to be created which firstly considers the roles stored in the Role Catalogue and associated permissions to forth coming changes that will necessitate defining new role, modify an existing role and removal/archiving of obsolete/unused roles. This detection mechanism can be considered as an iterative process which is carried out on a periodic basis or triggered due to a business event in a proactive manner at the time of business process change. In other words events that cause changes in organizational relationships require role memberships to be re-evaluated such that user access and privileges are in line with business policies; in doing do, the processes will feed back to Role Analysis and Engineering processes.

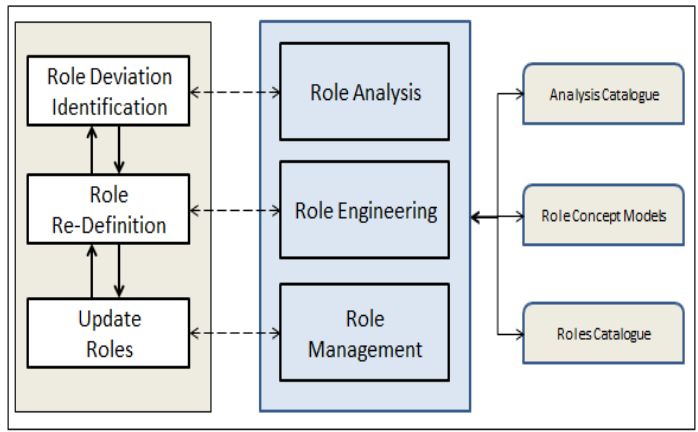

**Fig 14. Role Maintenance process interaction**

We can define this concept as shown in Fig 14. Role Deviation Identification upon initiation either on a periodic basis or upon event will consider the changes in the form of meta-data against existing users and Roles that are presently deployed. It is responsible for identifying the deviations from present setup including, conflicts or gaps or additional analytical information that can be studied by the respective Role administrators and business functional leads; this work may also involve traversing and reviewing the organizational structure change either bottoms-up or top down to validate these proposed changes. This process will work hand in hand with Roles Analysis process to ensure analysis is retained and updated in Analysis Catalogue.

Role Re-Definition will then be used to formalize the change in terms of approval, permissions, segregation of duties. It works closely with Roles Engineering which leverages defined techniques to manage exceptions, review changes to existing roles and or creating new roles, prototyping and performing validations with business partners until signoff is achieved. Roles Concept Model is then updated with these changes. Update Roles process will formally implement the change in Roles Catalogue and user information via Roles Management process. Any anomalies such as out of role exceptions, rights violations, duplicates found will require iteration back to prior step.

Potential exists for employing a form of rule based automated analysis to identify deviation which then feeds the information over to Roles Analysis process in a cyclical manner. Similarly rules can be written for roles defined in Roles Concept Model and Roles Catalogue to aid in the overall roles maintenance process becoming far simpler, efficient and streamlined while providing increased business agility.

## IV. Conclusion and Future Work

R12 EBS RBAC offers very powerful and flexible access control security than traditional responsibility. Even though RBAC security is more involved in setup; it offers easier administration and audit of users once a formal Roles Lifecycle Management approach is adopted as part of system administration. Roles Lifecycle Management requires strong business and technical processes, effective reviews and tight enforcement and integration across analysis, engineering, management and maintenance phases.

This paper represents a first detailed attempt at dissecting what a roles lifecycle management will entail to assist with Oracle R12 EBS and Fusion Financial Enterprise Planning Software; we have shown four tightly coupled processes that are integrated into a single comprehensive lifecycle each reflecting specific sub processes and outcomes. This work can be moved forward with more focus on Role Maintenance stage specifically in relation to metrics, identification and managing role changes.

## Acknowledgment

Special thanks for Dr Reza Samavir for his continued support and assistance with various discussions and thoughts towards focusing this particular research area.